**Title: Interactions Between Stably Rolling Leukocytes In Vivo**

**Authors:**

**Michael R. King, Aimee D. Ruscio, Michael B. Kim**

Department of Biomedical Engineering

University of Rochester

Rochester, NY 14642

**Ingrid H. Sarelius**

Department of Pharmacology & Physiology

University of Rochester

Rochester, NY 14642

**Abstract:**

We have characterized the two-dimensional spatial dependence of the hydrodynamic interactions between two adhesively rolling leukocytes in a live venule in the mouse cremaster muscle. Two rolling leukocytes were observed to slow each other down when rolling together in close proximity, due to mutual sheltering from the external blood flow in the vessel lumen. These results are in agreement with a previous study of leukocyte rolling interactions using carbohydrate-coated beads in a parallel-plate flow chamber and a detailed computer model of adhesion in a multicellular environment.

**Main text:**

The slow rolling motion of leukocytes (white blood cells) along the vessel wall in post-capillary venules is necessary step in the cellular immune reponse [1]. The dynamics of leukocyte rolling, as controlled by the kinetics of selectin-carbohydrate bond dissociation and other factors, has been mostly studied in flow chamber systems at very dilute concentrations of cells [2]. However, in vivo, leukocyte adhesion with the vessel wall occurs in the presence of a dense suspension of blood cells. The effects of the surrounding suspension of cells on leukocyte adhesion to the endothelium have not been fully elucidated. In previous studies we have begun to determine these effects by formulating a computational model of cell-surface interactions under flow in a multicellular environment, that fuses a stochastic simulation of receptor-ligand binding with a rigorous boundary elements calculation of suspension flow at zero Reynolds number [3]. The multiparticle adhesive dynamics (MAD) simulation agrees quite well with flow chamber experiments using carbohydrate-coated beads rolling on a purified P-selectin surface [3-5]. One of the main results of this work has been to demonstrate that a leukocyte will roll at a lower velocity when in close proximity to another rolling leukocyte. Ref. 3 shows the spatial dependence of this effect in both the flow ($x$) and transverse ($y$) directions, since in the flow chamber experiments the entire substrate surface is viewed from below. Recently, we have shown that some of these hydrodynamic effects may be observed in live hamster microvessels [6]. By taking midplane cross-sections of straight, unbranched vessels to produce a side-view of leukocyte-wall interactions, we were able to show a correlation between leukocyte rolling velocity and the distance between rolling leukocytes in the flow direction only. A decrease in rolling velocity with decreasing cell-cell separation was observed, in agreement with a 1/$r$ hydrodynamic scaling and in qualitative agreement with previous microsphere and computational results.

In this Brief Report we describe new experiments in anesthetized mice, in which the upper wall of large (~40 μm diameter) post-capillary venules was focused on. This produced a top-view of an approximately planar surface near the center of the region, and the relative $x$-$y$ positions of rolling leukocytes can be measured. Thus, the fully two-dimensional spatial relationship of leukocyte-leukocyte hydrodynamic interactions as predicted in Ref. 3 could be tested in a live animal vessel. Indeed, we show that the two-dimensional map of rolling velocities of pairs of rolling leukocytes is in good qualitative agreement with our previous report.

All protocols were approved by the Institutional Animal Care and Use Committee of the University of Rochester. Male C57BL/6J mice weighing 28 to 32g and older than 8 weeks were used for these experiments. Animals were anesthetized with an initial dose of sodium pentobarbital (75mg/kg ip) and maintained with supplemental i.v. doses as needed. Anesthetic level was monitored by observing withdrawal reflexes initiated by toe or tail pinch. Body temperature was maintained by placing the animal on a glass heating coil connected to a water circulator set at 37°C. The animal was tracheotomized to establish a patent airway and a catheter (PE tubing) was placed in the right jugular vein for delivery of anesthetic. The right cremaster muscle was prepared for in situ microscopy as previously described [7]. The tissue was superfused with a bicarbonate buffered salt solution throughout the surgical preparation and subsequent observation. The superfusate was maintained at physiological temperature (36±0.5°C) and equilibrated with gas containing 5% $CO_2$ and 95% $N_2$ to maintain physiological pH (7.40±0.05). At the end of the experiment, the animal was euthanized by i.v. injection of a lethal dose of anesthetic. The upper wall of post capillary venules of ~40 μm in diameter was observed using an Olympus BX50WI microscope equipped with a water immersion objective (Olympus LumPlanflo 40X, 0.80 NA).

Following the surgical procedure to exteriorize the mouse cremaster muscle tissue, mild inflammation consistent with P-selectin-mediated adhesion was observed. Figure 1 shows digitized video images of pairs of rolling leukocytes in a typical vessel. Only pairs of rolling leukocytes far from other cells were used in our measurements, to focus on binary leukocyte-leukocyte interactions. The center-to-center separation distance in the flow ($x$) and transverse ($y$) directions was measured for each pair, as well as the translational rolling velocity of each cell from successive video frames. Time intervals of 0.2-1 s were used to determine cell velocities, with each cell producing ~5 velocity measurements, for a total of 1500 values.

Figure 2 shows the two-dimensional map of leukocyte rolling velocity for the 150 cell pairs tracked in the mouse venule. The wall shear rate in this vessel was determined to be 220 $s^{-1}$, by tracking the motion of fluorescent tracer particles as described previously [8]. Velocities are plotted as function of the separation distance between the two cells that comprise each cell pair. As can be seen from the blue region of low velocity, when two cells are close to one another the average rolling velocity is reduced by a factor of two as compared to the average velocity for isolated cells (30 vs. 60 μm/s). Furthermore, this effect is more pronounced when the cells are aligned in the direction of flow, as opposed to two cells rolling side-by-side which translate at a somewhat higher velocity of 40 μm/s on average. For cell pairs separated by 4 cell radii (16 μm), the rolling velocity is equal to 55 μm/s, or ~90% of the isolated cell velocity.

The spatial dependence of hydrodynamic interactions between rolling leukocytes, as indicated in Figure 2, is in complete agreement with our previous work using 10.8 μm sialyl Lewis$^X$-coated microspheres rolling on a P-selectin surface, as well as the detailed MAD simulation of concentrated cell rolling. This phenomenon is due to each spherical leukocyte reducing the drag force on each other, by creating a disturbance flow that decays as $1/r$ from each tethered cell. This Brief Report thus represents the first demonstration that the two-dimensional spatial dependence of leukocyte interactions as predicted by MAD is in fact measurable in real in vivo microvessels. The biological implication of these results is that as leukocytes accumulate at a site of inflammation, there will be a collective physical effect acting to slow their progress through the circulation. Future work with these experimental and computational models of cellular interactions will fully determine the effects that the multicellular nature of whole blood has on leukocyte recruitment and adhesion.

**Acnowledgement:**

This work was supported by NIH Grant HL018208.

**Figure Captions:**

**Figure 1.** Digitized video images of leukocytes rolling on the upper wall of a 40 μm venule in mouse cremaster muscle. A. and C. show pairs of cells with center-to-center separation distances of (Δx,Δy) = (16,5.5) and (1,10) μm, respectively. For comparison, the diameter of each leukocyte is about 8 μm. B. defines the coordinate system.

**Figure 2.** Map of rolling velocities of pairs of leukocytes in a live post-capillary venule of diameter 40 μm in an anesthetized mouse. The average rolling velocity is plotted versus the center-to-center separation distance between the two cells in the flow ($x$) and transverse ($y$) directions. The red box in the upper right corner of the graph represents the average rolling velocity of isolated leukocytes ($N = 8$) in the same vessel.

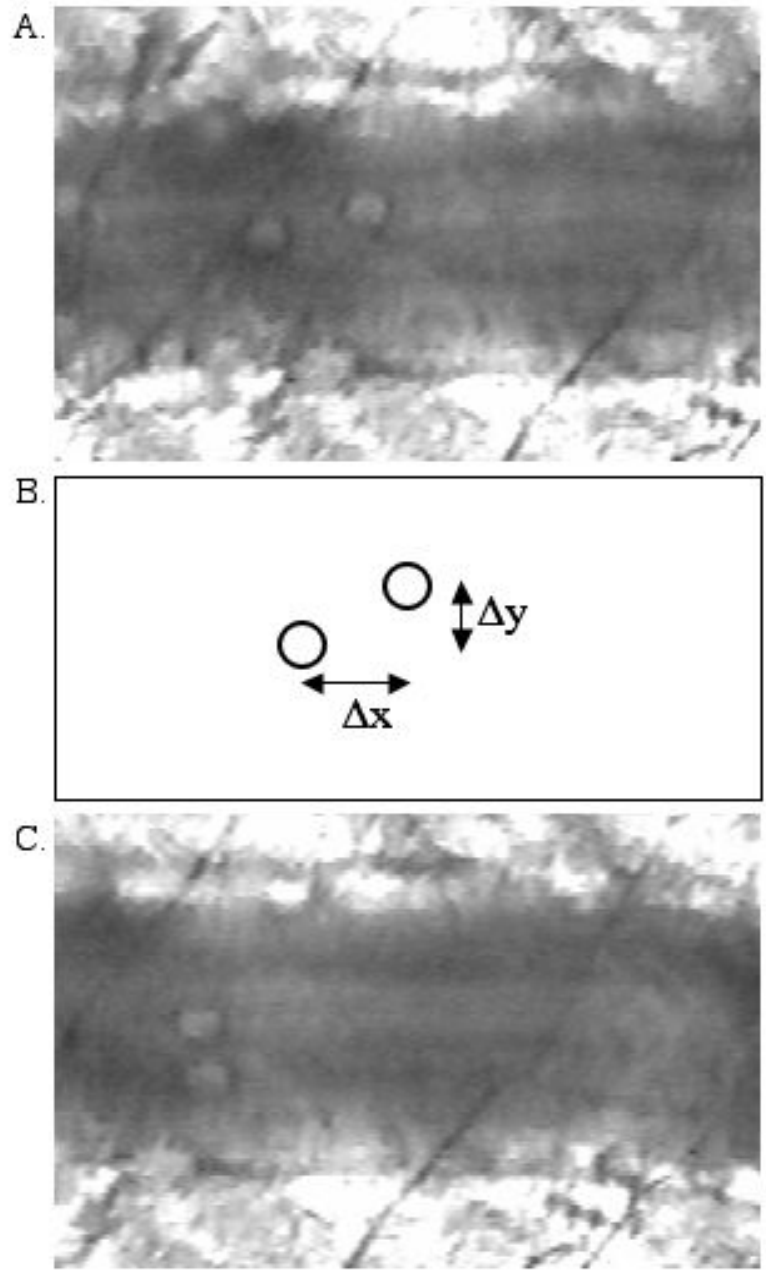

**Figure 1.**

M.R. King et al.

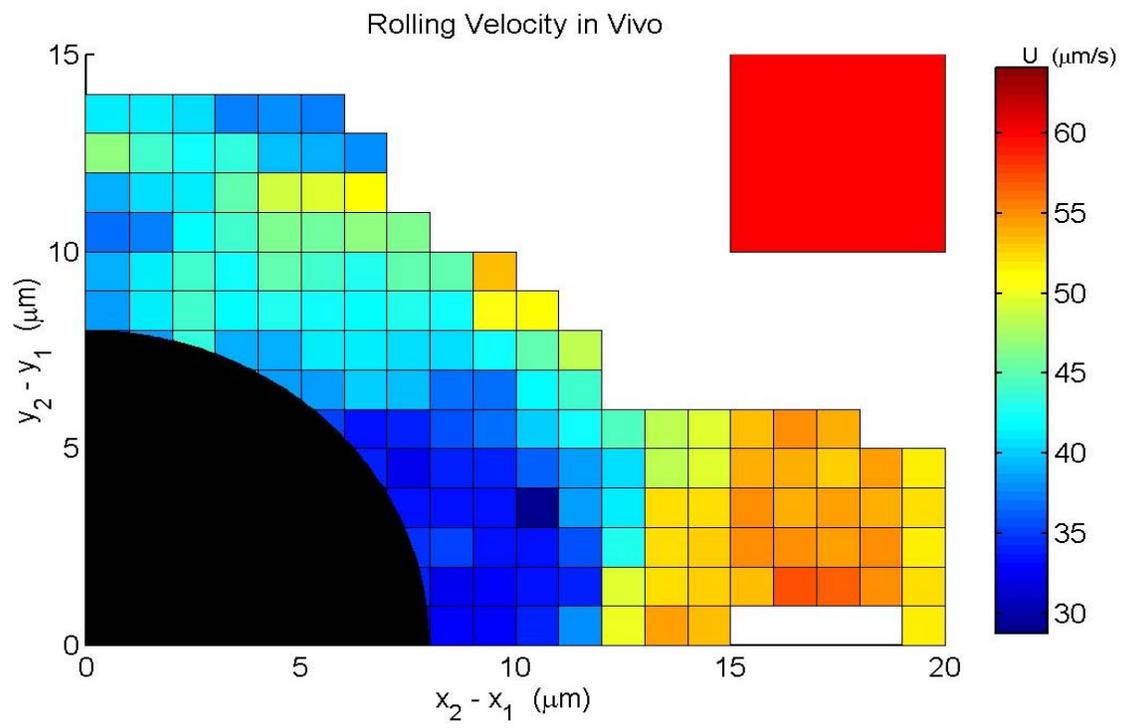

**Figure 2.**

M.R. King et al.